\documentclass{appolb}
\usepackage{graphicx}
\usepackage{times}
\usepackage[english]{babel}
\usepackage{amsmath}
\usepackage[latin1]{inputenc}
\usepackage[T1]{fontenc}
\usepackage{graphicx}
\usepackage{epsfig}
\usepackage{rotating}
\usepackage{subfigure}
\usepackage{tikz}
\usepackage[subfigure]{ccaption}
\usepackage{mathtools,cancel}

\def\ii{\'{\i}}

\begin{document}
\title{Observables in the 3 Flavor PNJL Model and their Relation to Eight Quark Interactions.%
\thanks{Presented at the workshop "Excited QCD 2012",
06-12 May 2012 in Peniche, Portugal. Work supported by FCT, 
CERN/FP/116334/2010, QREN, UE/FEDER through COMPETE. 
Part of the EU Research Infrastructure Integrating Activity 
Study of Strongly Interacting Matter (HadronPhysics3) under the 7th Framework 
Programme of EU: Grant Agreement No. 283286.}%
}
\author{B. Hiller\footnote{Speaker}, J. Moreira, A. A. Osipov, A. H. Blin
\address{Departamento de F\ii sica, CFC, Faculdade de Ci\^encias e Tecnologia da Universidade de Coimbra, P-3004-516 Coimbra, Portugal}
}
\maketitle
\begin{abstract} 
Several relevant thermodynamic observables obtained within the (2+1) flavor and spin zero NJL and PNJL models with inclusion of the 't Hooft determinant and $8q$ interactions are compared with lattice-QCD (lQCD) results. In the case that a small ratio $R=\frac{\mu_B}{T_c}\sim 3$ at the critical end point (CEP) associated with the hadron gas to quark-gluon plasma transition is considered, combined with fits to the lQCD data of the trace anomaly \cite{Bazavov:2009}, subtracted light quark condensate \cite{Bazavov:2009} and continuum extrapolated data of the light quark chiral condensate \cite{Bazavov:2012}, a reasonable description for the PNJL model is obtained with a strength  $g_1\sim 5...6 \times 10^3$ GeV$^{-8}$ of the $8q$ interactions. The dependence on the further model parameters is discussed.  
\end{abstract}
  
\vspace{0.5cm}

In recent years the role of effective chiral Lagrangians  has grown as an important indicator of the  order and universality class of phase transitions, as well as of the nature and location of the related CEP that may occur for the ground state of QCD in presence of external parameters, such as finite temperature $T$, baryonic chemical potential $\mu_B$, magnetic field $B$ \cite{Pawlowski:2011}. In parallel, lQCD advances at zero and moderate chemical potential with masses approaching the physical values of the light quarks \cite{Bazavov:2012} and pion mass \cite{Ratti:2012}, strongly indicate at a crossover transition from the hadronic to the quark-gluon phase at finite $T$ and $\mu_B=0$. Combining lQCD and chemical freeze-out data from relativistic heavy-ion collision facilities, the location of the CEP is presently conjectured to eventually occur at $R=\frac{\mu_B}{T_c}\sim 2$ and $\frac{T}{T_c}\sim 1$, \cite{Stephanov:2004},\cite{Gupta:2011}.

We consider the $SU(3)$ flavor and spin-0 Nambu-Jona-Lasinio model (NJL) \cite{Nambu:1961} with inclusion of the $U(1)_A$ breaking 't Hooft flavor determinant \cite{Hooft:1976}-\cite{Reinhardt:1988} and eight quark ($8q$) interactions \cite{Osipov:2006b},\cite{Osipov:2007a} (of which there exist two types, one of them violationg the OZI rule, with strength $g_1$), and extend it to include the Polyakov loop (PNJL) \cite{Fukushima:2004}-\cite{Moreira:2011}. The $8q$ have been firstly introduced to stabilize the effective potential of the model \cite{Osipov:2006b}.  Their role turned out to be of significant importance in the behavior of model observables in presence of external parameters \cite{Osipov:2007b}-\cite{Hiller:2010},\cite{Kashiwa:2008a}-\cite{Gatto}. Of particular interest is that the $8q$ coupling strengths $g_1$ can be varied in tune with the $4q$ interaction strength $G$ without changing the vacuum condensates and low energy meson spectra, except for the $\sigma$-meson mass $m_\sigma$ which decreases with increasing $g_1$. Fits to the low lying pseudoscalar and scalar meson spectra yield $m_\sigma\sim 560$ MeV for $g_1=6000$ GeV$^{-8}$ and  $m_\sigma\sim 690$ MeV for $g_1=1500$ GeV$^{-8}$ \cite{Osipov:2007a}. 
In the $\mu,T$ plane (where $\mu=\frac{\mu_B}{3}$) the $g_1,G$ interplay gives rise to a line of CEP, starting from the regime of large ratios $R\sim 20$ (NJL) and $R\sim 10$ (PNJL) in the case of weak $8q$ coupling $g_1$, to small ratios for strong $g_1$. In the first case the chiral condensate is related with spontaneous symmetry breaking $(SSB)$ driven by $4q$ interactions, in the second scenario $SSB$ is induced by the $6q$ 't Hooft strength  \cite{Osipov:2007a},\cite{Osipov:2008},\cite{Hiller:2010}. This continuous set of CEP is particular to the $8q$ extension of the model.
However a correlation between $m_\sigma$ and the location of the CEP is also observed in the (2+1) - flavor quark-meson Lagrangian, where besides the 't Hooft term, a quartic mesonic contribution is present
\cite{Schaefer:2010},\cite{Chatterjee},  thus bearing a resemblance to the semi-bosonized version of the $8q$ NJL Lagrangian \cite{Osipov:2007a}. In order to restrict the $g_1$ values one may: i) calculate decays and scattering in the vacuum, which are expected to narrow the choice and ii) compare with available lQCD data at finite $T$ and moderate $\mu$. In the present study we try to explore the second option. For the PNJL case  an extra uncertainty arises due to the parameters related with the choice of Polyakov potential ${\cal U_{P}}$. In particular the $T_0$ parameter of \cite{Ratti:2006},\cite{Roessner:2007} has a sizeable effect on the  transition temperature. 
First we show in Fig.~1  the CEP lines in a $(\mu,T)$ vs. $g_1$ diagram. The PNJL model (solid lines) enhances the effect of pushing $R$ to small values as functios of $g_1$ in comparison with the NJL case (dashed lines).
 The crossing of the $CEP(T)$ and $CEP(\mu)$ lines, ( yielding $R=3$), is reached for the PNJL at $g_1\sim 6.4 \times 10^3$GeV$^{-8}$, for the choice $T_0=190$ MeV, whereas it occurs for the NJL only at a much larger value, $g_1\sim 8.4\times 10^3$GeV$^{-8}$ (we remind that with increasing $g_1$ the crossover becomes sharper  and eventually gives rise to a first order transition at $\mu=0$, which happens at $g_1\sim 9\times 10^3$GeV$^{-8}$ in the NJL case). Changing $T_0$, the $CEP(T)$ is shifted up (down) with increasing (decreasing) $T_0$ (see caption of Fig.1), while $CEP(\mu)$ remains sensibly unaltered. 
In Fig. 2 the chiral condensates and dressed Polyakov loop for $u,s$ quarks are shown for the NJL as function of $T$ for $\mu=0$ and with varying strength $g_1$ (see caption). For $g_1 10^{-3}=1;5;6.5;8$ GeV$^{-8}$ the transition temperatures $T_t$ defined at the corresponding inflection points of the curves are $T_t=192;163;147;135$ MeV for the u-condensate, $T_t=197;163;150;135$ MeV for the u-quark dressed Polyakov loop, $T_t=197;160;147;135$ MeV at the first inflection point of s-quark condensate, $T_t=270;240;235;225$ MeV at its 2nd inlection point, $T_t= - ;166;150;135$ MeV at 1st inflection point of the dressed s-quark Polyakov loop and $T_t=270;240;235;225$ MeV at its 2nd inflection point. The 1st set of inflection points in the case of the s-quark condensate and dressed Polyakov loop occur due to the gap equations that correlate the $u$ and $s$ variables, yielding similar $T_t$ for the $u$ and $s$ observables. The 2nd inflection point occurs at temperatures $T_t$ larger by $\sim 80$ MeV. A similar pattern is observed for the PNJL model in Fig. 3, the second inflection points occur at roughly $55$ MeV higher $T_t$ values. Visually these 2nd inflection points can barely be detected, the transition is very slow and smooth. This behavior can be traced back to the fact that for large $T$ the $s$-quark constituent quark mass approaches asymptotically its current quark mass value, which is much larger than for the $u$-quark.\footnote{We calculate the thermodynamic potential with the prescription of \cite{Hiller:2010},\cite{Moreira:2011} where we show that it leads to the correct large $T$ asymptotic behavior for the quark masses (condensates), traced Polyakov loop and number of degrees of freedom.}  
 It is a disputable matter which temperature should be taken to characterize the transition for the $s$-quark in these observables. A calculation of the chiral and quark number susceptibilities associated with the s-quark in the NJL model display only one peak characterizing the transition temperature.
In Fig. 4 one sees however that in the PNJL case two peaks can occur again for the s-quark chiral susceptibility. 
Fig. 5 (a) shows the trace anomaly calculated for $g_1=6000$ GeV$^{-8}$, for various values of the parameter $T_0$ in comparison with lattice data. In Fig. 5 (b) the subtracted condensate $\Delta_{ls}$  is shown for several values for $g_1$, calculated with $T_0=.19$ GeV, and compared to lQCD. In Fig. 6 the light chiral condensate is compared with lQCD data extrapolated to the continuum limit \cite{Bazavov:2012} for different values of $g_1$ and with ${\cal U_{P}}$ of \cite{Ratti:2006} for the cases $T_0=.15$ GeV (left) and $T_0=.19$ GeV (right).




\begin{figure}[htb]
\centerline{\includegraphics[height=4cm,width=6cm]{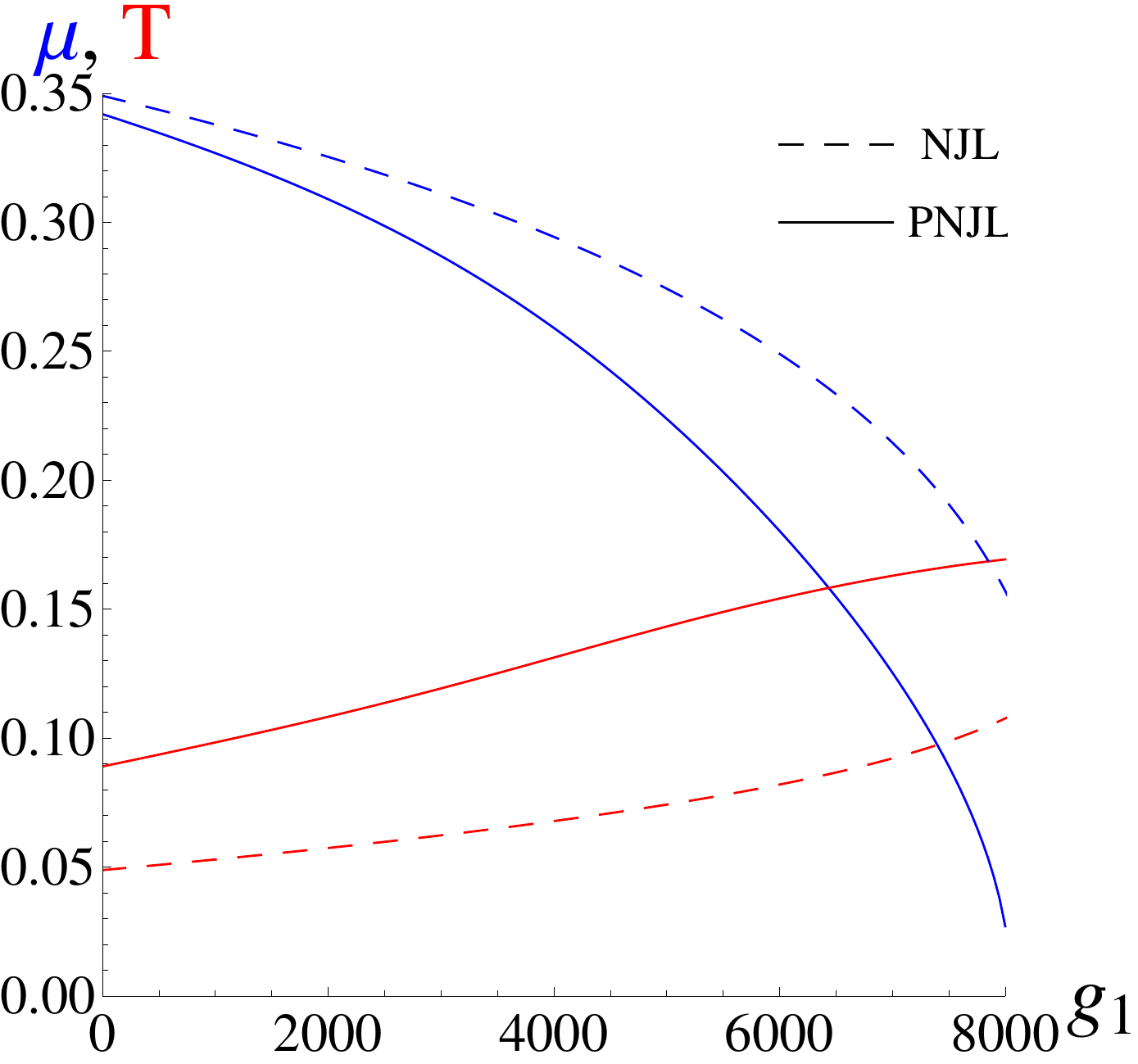}}
\caption{\footnotesize Pairs $(T,\mu)$ of CEP as function of the $8q$ interaction $g_1$. Positive slope lines (red online) show T-dependence, negative slope lines (blue online) show $\mu$ dependence.  All model parameters fixed as in \cite{Hiller:2010}, except for $g_1,G$. PNJL potential from \cite{Ratti:2006}. Intersection ($R=3$) of PNJL curves (solid lines) occur at ($\mu=T=158;167;188$ MeV) with $g_1=6436;6251;6127$ GeV$^{-8}$ for $T_0=190;210;270$ respectively (shown only for $T_0=190$ MeV). Intersection of NJL curves (dashed lines) at $\mu=T=117$  with $g_1=8372$ GeV$^{-8}$.}
\end{figure}


\begin{figure}[htb]
\begin{minipage}[c]{0.6\linewidth}
 \centering
 \includegraphics[height= 0.15 \textheight,width=0.6 \textwidth]{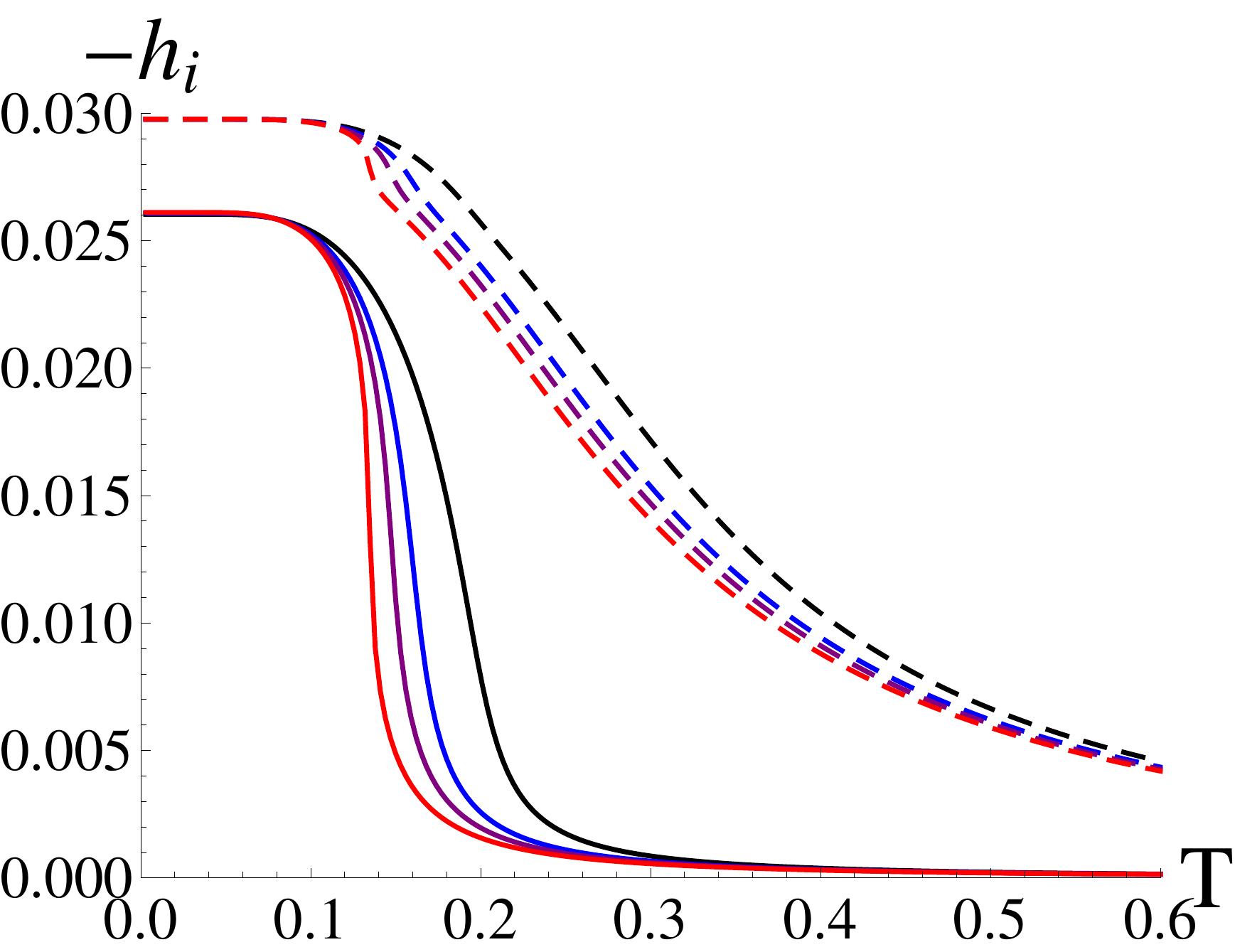}\\
 \includegraphics[height= 0.15 \textheight,width=0.6 \textwidth]{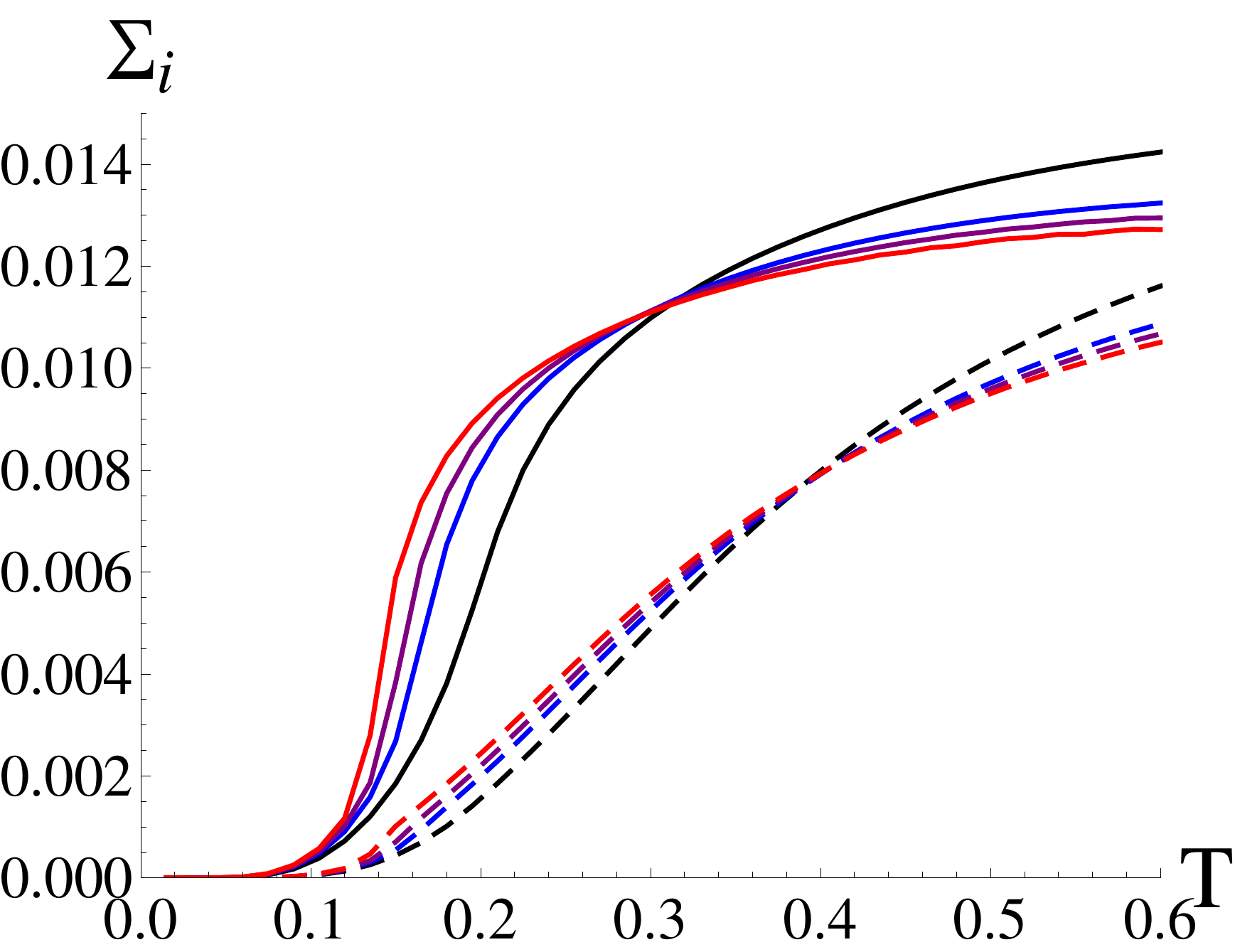}
\caption{\footnotesize The chiral condensates $=h_i/2$, $i=u,s$ and the dressed Polyakov loop $\Sigma^i$ as functions of T for NJL;  solid lines for light quarks, dashed for the strange quark. Up to down curves in upper panel: $g_1\times 10^{-3}=1;5;6.5;8$ GeV$^{-8}$, corresponding to black, blue, violet, red (color online). 
}   
\end{minipage}
\hspace{0.5cm} 
\begin{minipage}[c]{0.6\linewidth}
  \centering
  \includegraphics[height= 0.15 \textheight,width=0.6 \textwidth]{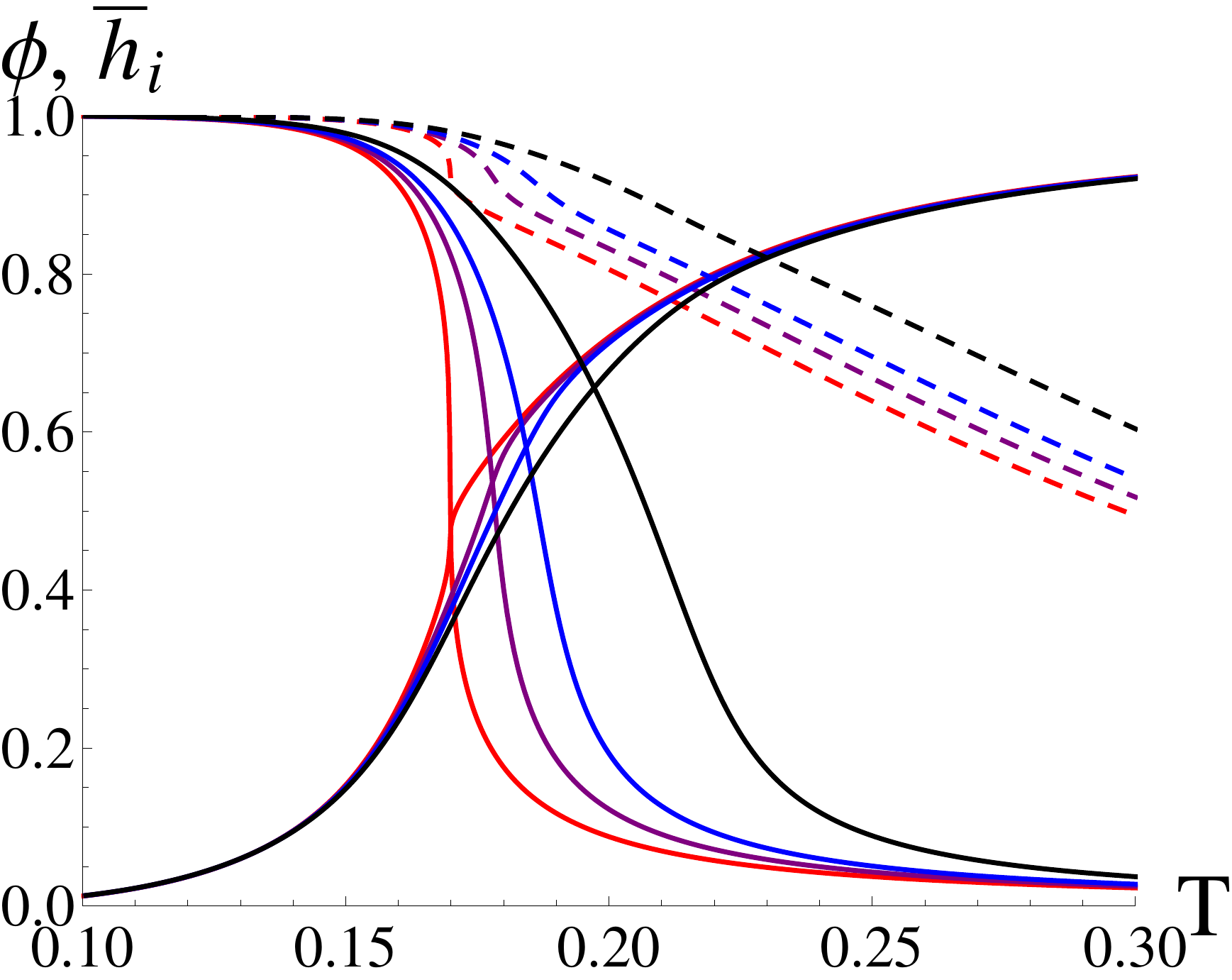}\\
  \includegraphics[height= 0.15 \textheight,width=0.6 \textwidth]{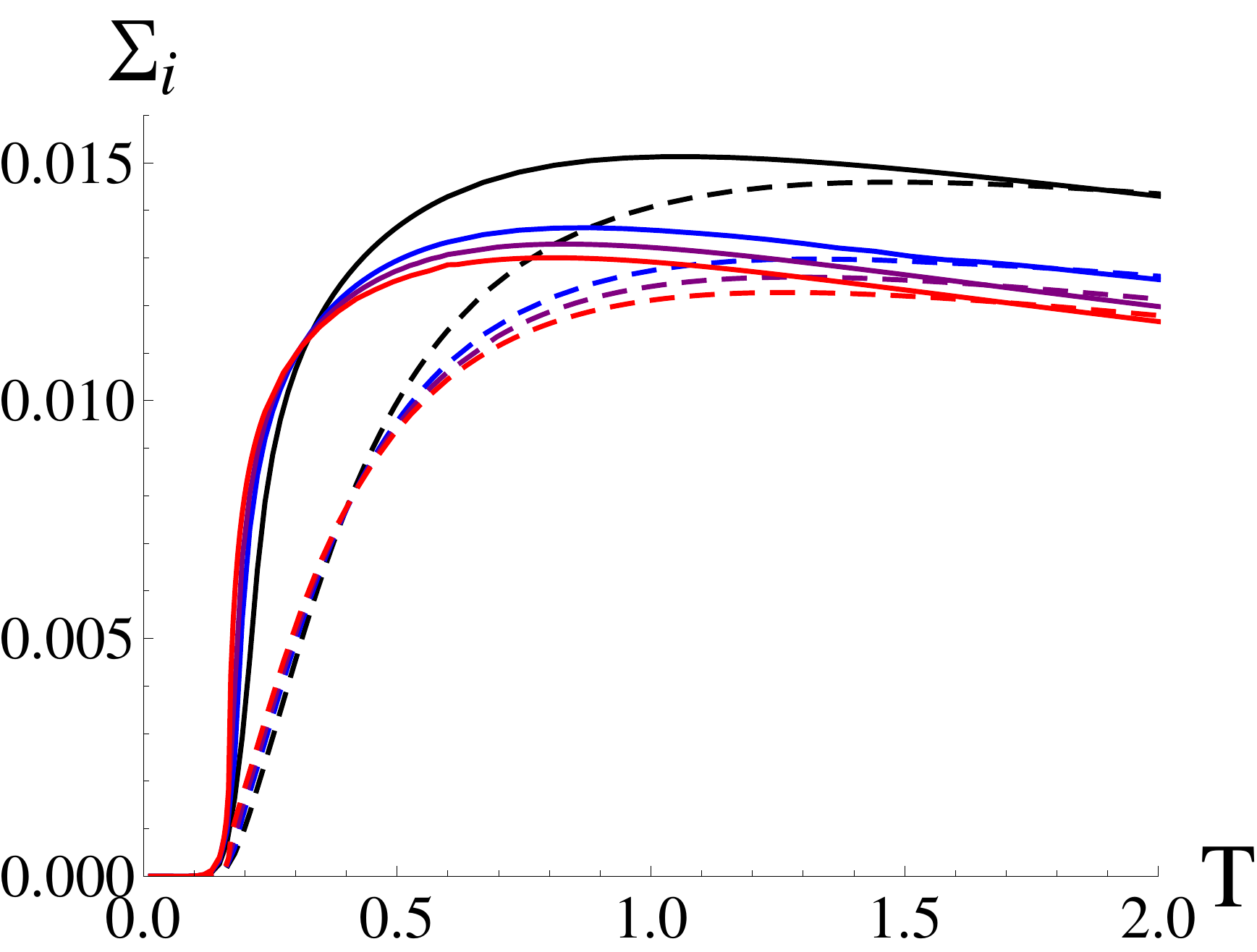}
\caption{\footnotesize The same as in Fig.1 for the PNJL model. In upper panel $\phi$ (curves growing with $T$) stand for the Polyakov loop and $g_1$ strength reverts order compared to the chiral condensate. }
\end{minipage}
\end{figure}







\begin{figure}[htb]
\begin{center}
\subfigure[]{\includegraphics[height= 0.133 \textheight]{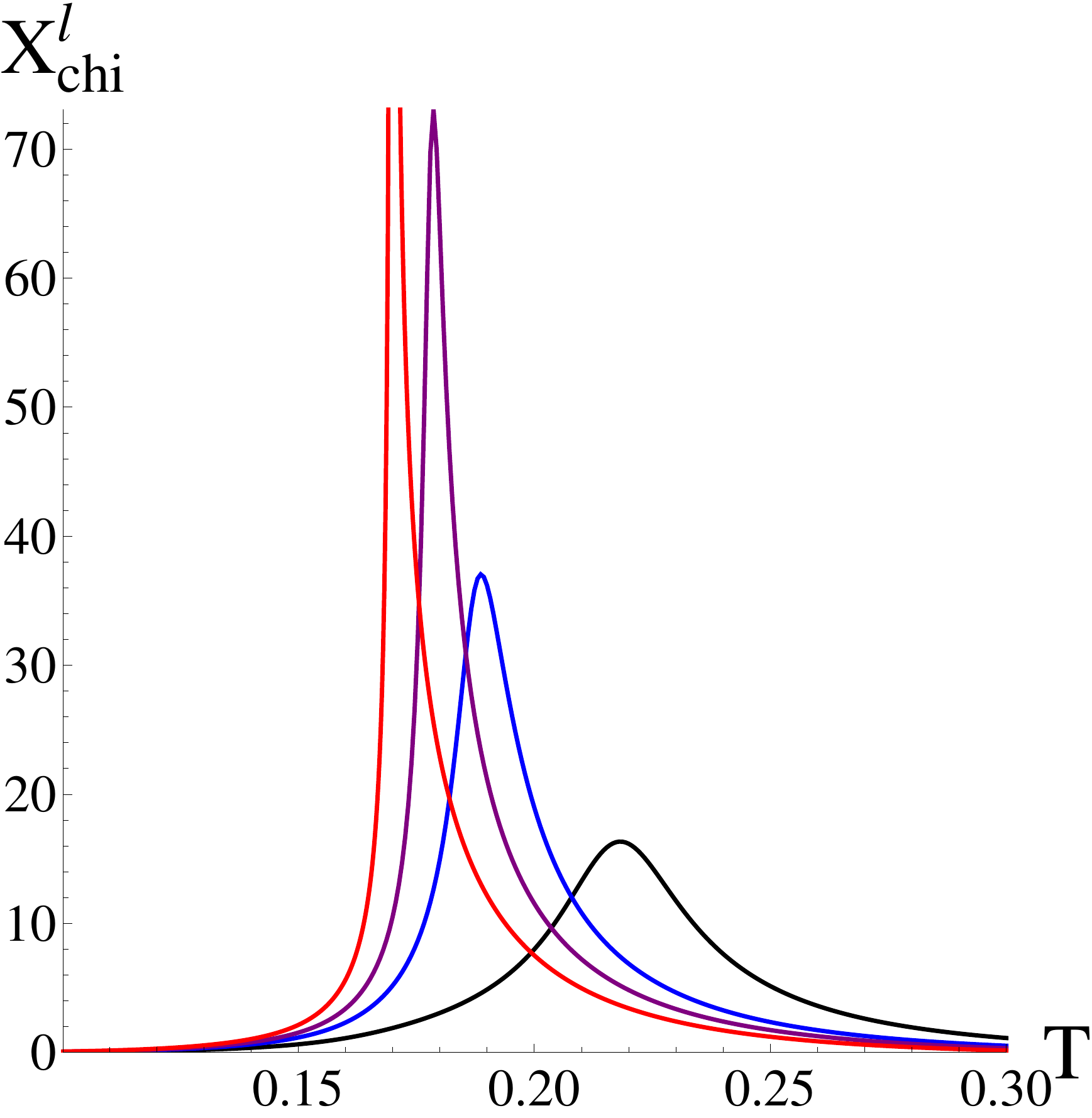}}
\subfigure[]{\includegraphics[height= 0.133 \textheight]{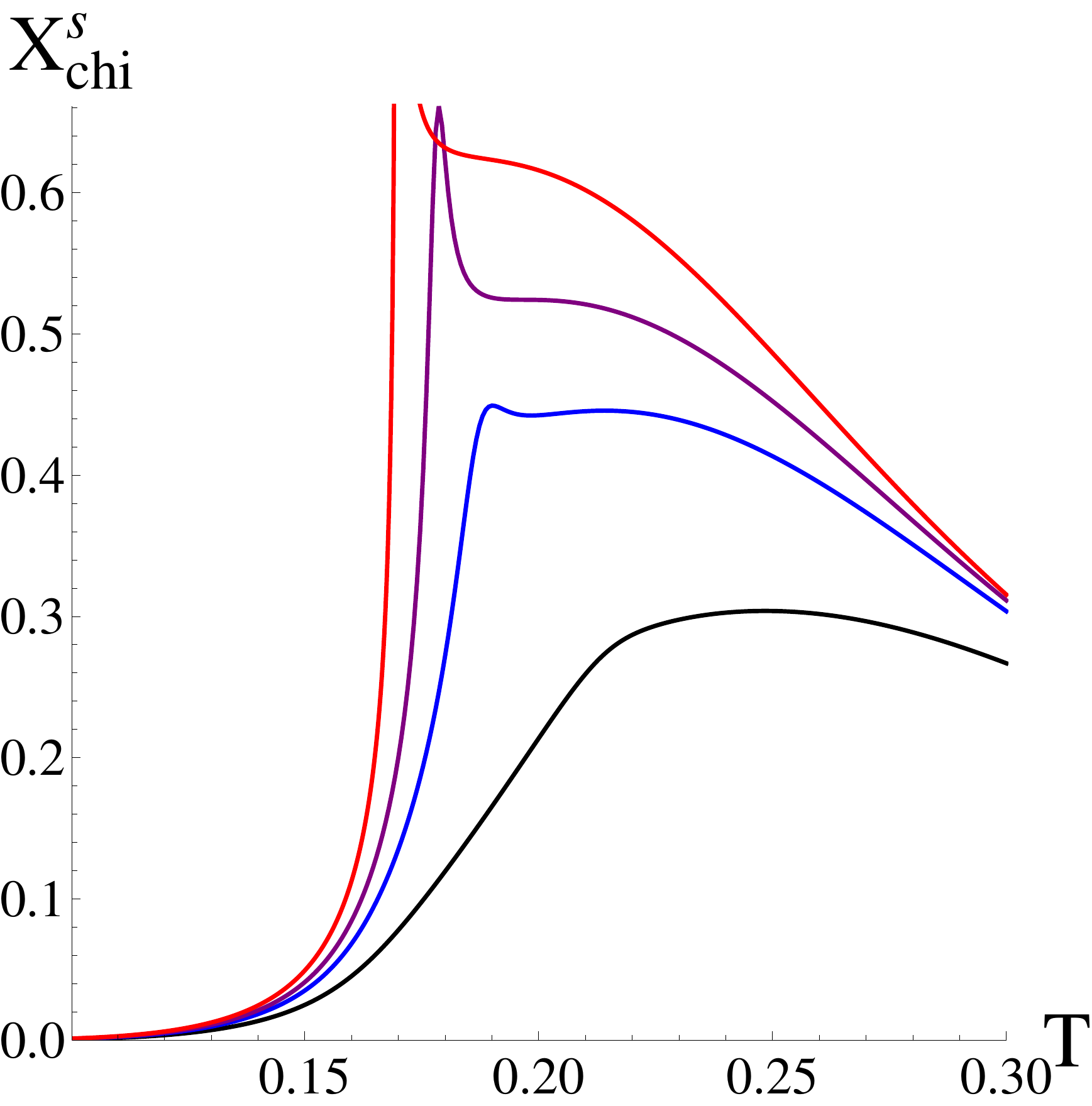}}
\subfigure[]{\includegraphics[height= 0.133 \textheight]{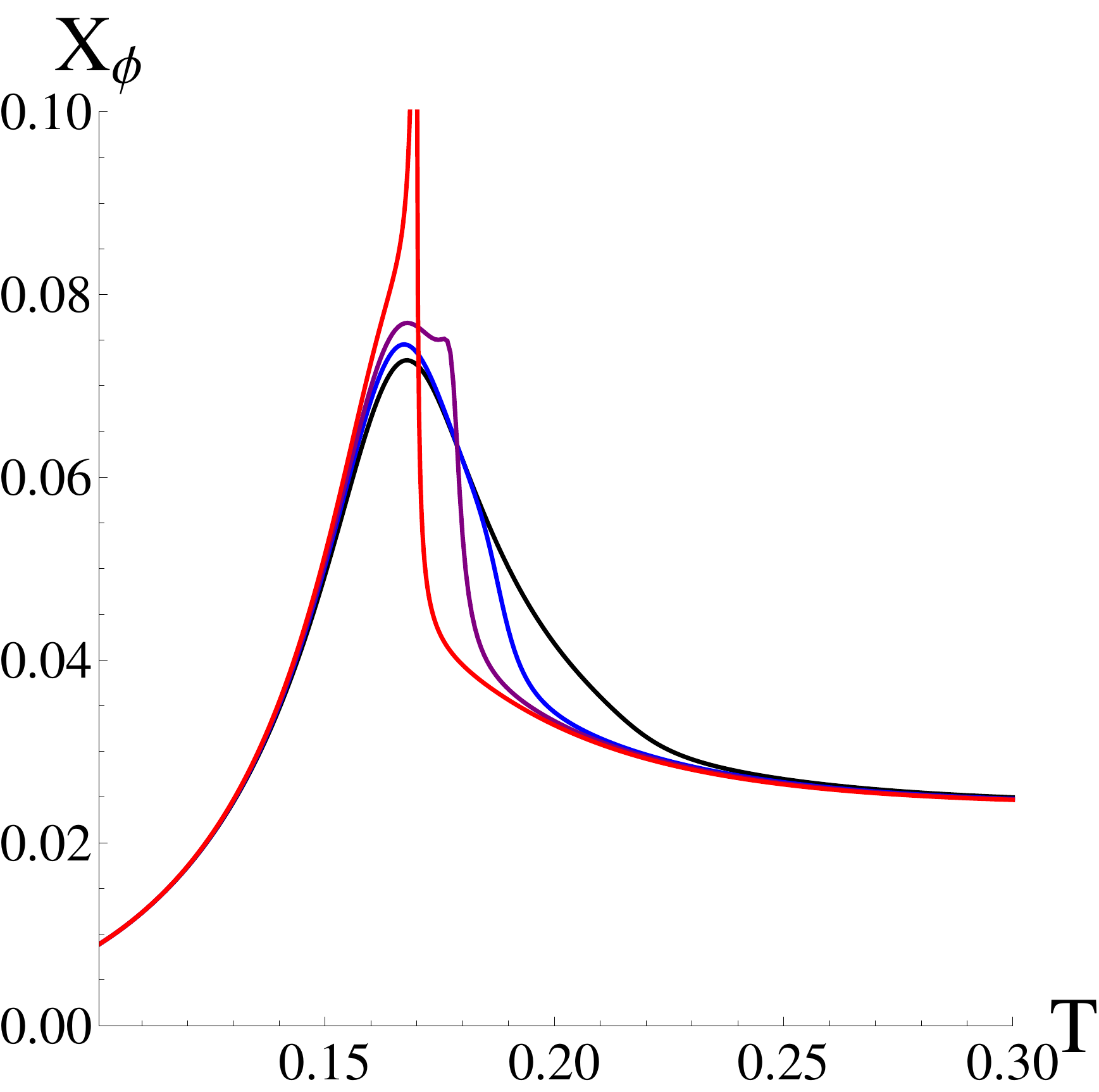}}
\subfigure[]{\includegraphics[height= 0.133 \textheight]{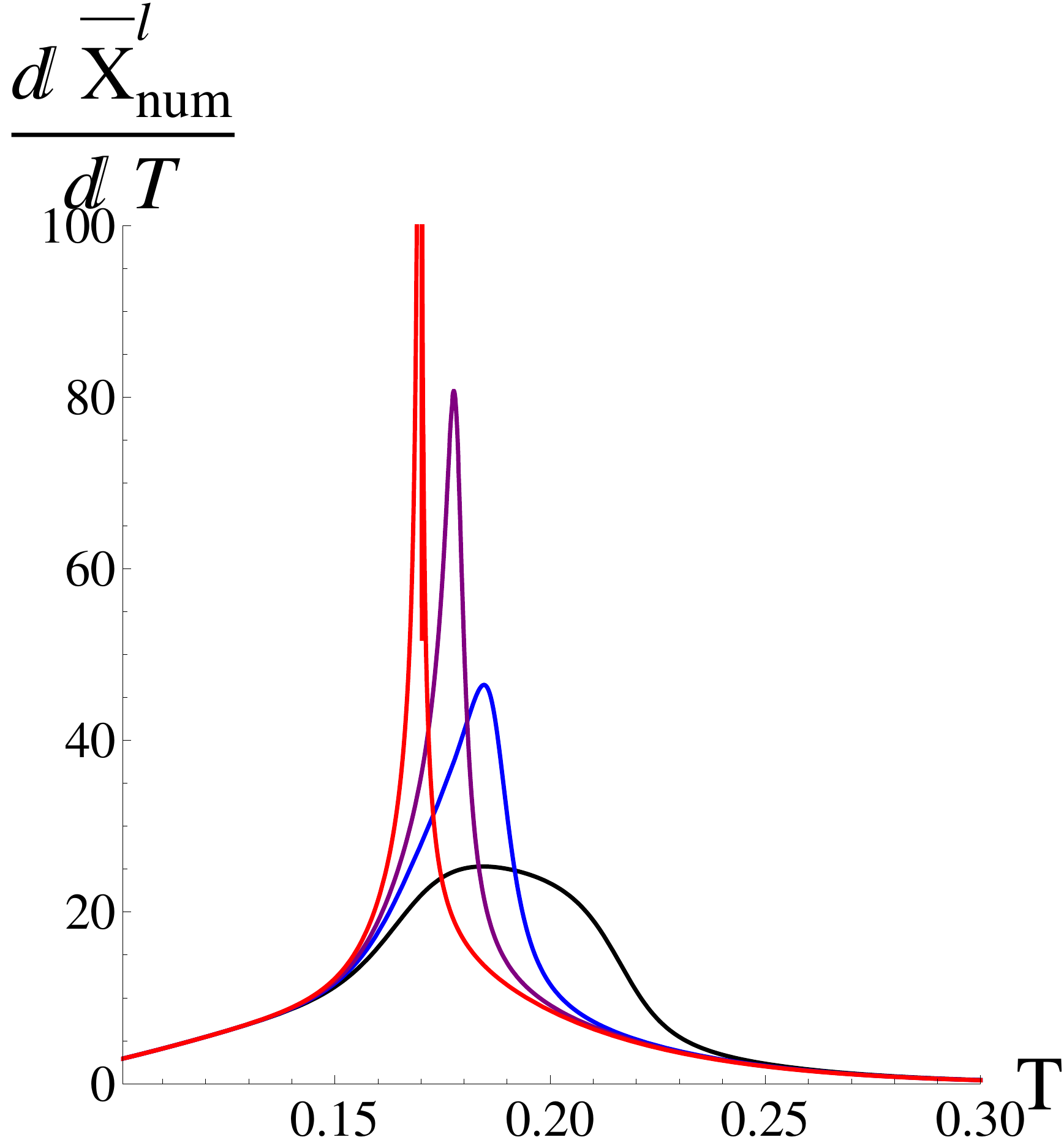}}
\subfigure[]{\includegraphics[height= 0.133 \textheight]{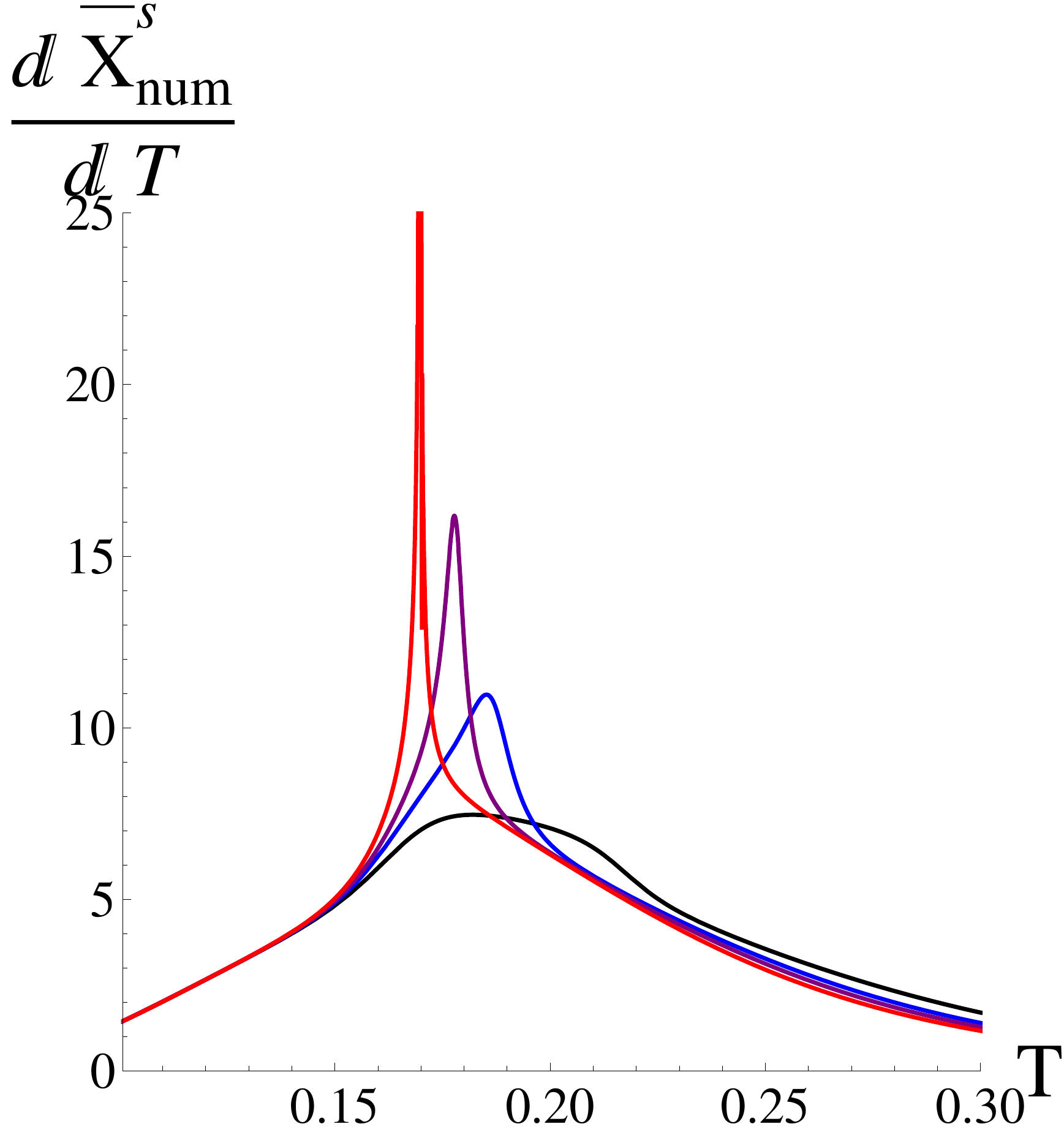}}
\caption{\footnotesize {Susceptibilities for PNJL, ${\cal U_{P}}$ of \cite{Ratti:2006} and $T_0=190$MeV. a) light quark chiral susceptibilites, (b) $s$ quark  chiral susceptibilities, (c) Polyakov loop susceptibility, (d) quark number (for $u$ quark) and  (e) quark number (for $s$-quark) susceptibilities. The peaks get more pronounced with increasing $g_1$. Color code as in Fig.2.}}
\end{center}
\end{figure}
From these comparisons we conclude (i) that the smaller the ratio $R=\frac{\mu_B}{T_c}$ related with the CEP location, the larger the $8q$ interaction strength $g_1$ must be chosen; a sizeable dependence on the $T_0$ parameter of the Polyakov potentials can induce shifts of the order of several tens of MeV in $T_c$ (Fig. 1). For $R=3$ we get $g_1$ of the order $6000$ GeV$^{-8}$ and $T_c=158-188$ MeV for the range $T_0=190-270$ MeV. (ii) Besides the $8q$ strength, the Polyakov loop plays also a substantial role in decreasing the ratio $R$. (iii) The  observables calculated at $\mu=0$ related with the light quarks, chiral condensates, traced Polyakov loop and dressed Polyakov loop (Fig. 3), chiral and quark number susceptibilities (Fig. 4 (a), (d)), as well as the s-quark number susceptibility (Fig. 4 (e)) and Polyakov loop susceptibility (Fig. 4 (c)) yield a crossover temperature $T_t\sim 179$ MeV for $g_1=6000$ GeV$^{-8}$ and $T_0=.19$ GeV. (iv) Some of the s-quark observables show two possible transition temperatures, Fig.2,3,4(b), the first close to the u-quark transition, the second about 50 MeV higher for the PNJL model. (v) The best fit to the trace anomaly is for $g_1=6000$ GeV$^{-8}$ at $T_0=.21$GeV (Fig. 5 (a)) and for the observable $\Delta_{ls}$ we obtain a reasonable fit with $g_1=5...6\times 10^3$  GeV$^{-8}$ and $T_0=.19$ GeV (Fig. 5 (b)). (vi) The peak positions and heights of the continuum extrapolated light quark chiral susceptibility vary considerably  (Fig. 6). This big spread  allows to accomodate a large range of $g_1$ values, whose peak positions in turn depend also on the choice of the $T_0$ parameter. The value $g_1\sim 5\times 10^3$GeV$^{-8}$ is eventually the best choice if one takes the height of the peak also into consideration.              
\begin{figure}[htb]
\begin{center}
\subfigure[]{\includegraphics[height= 0.2 \textheight,width=0.4 \textwidth]{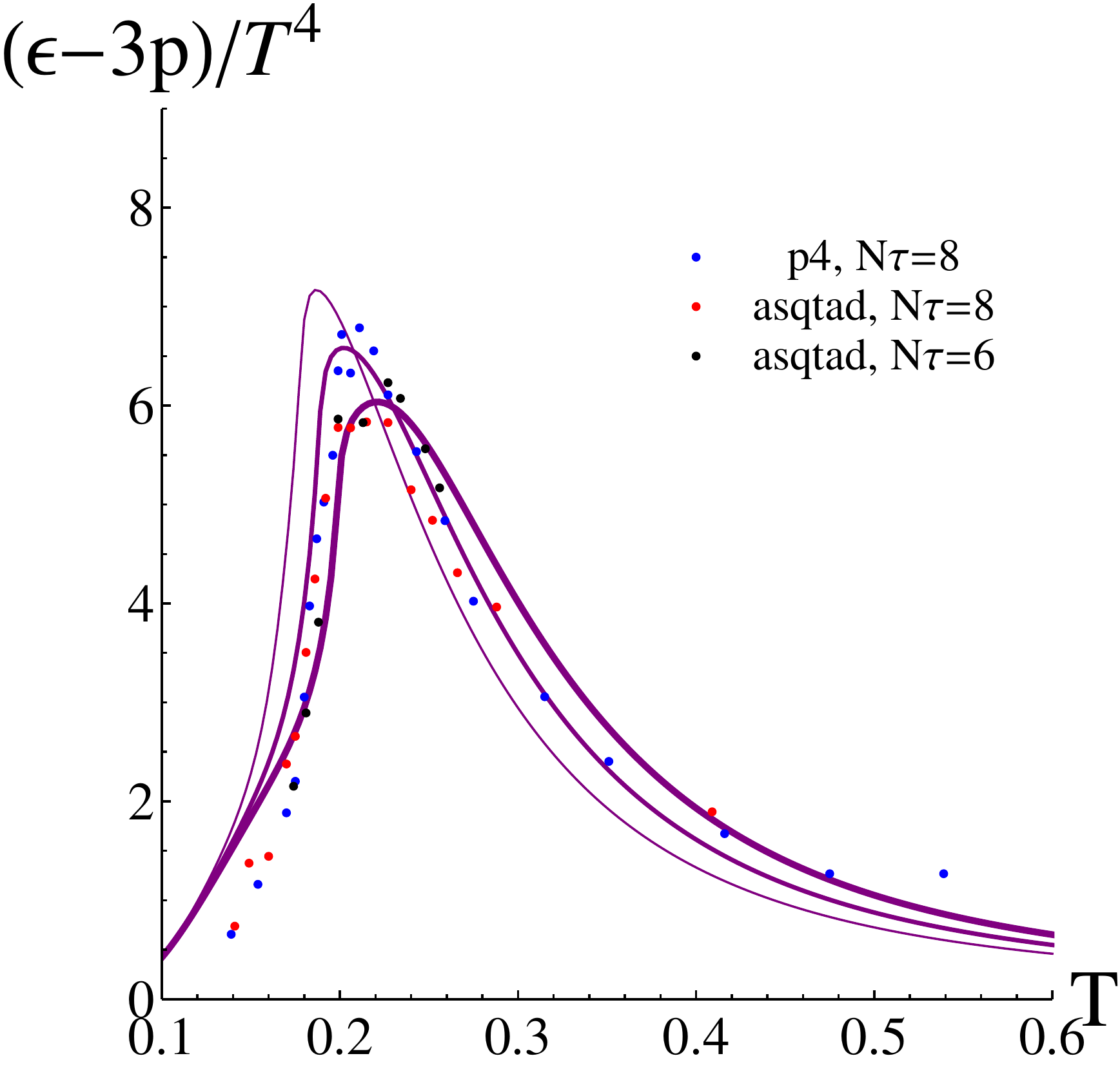}} \hspace{0.2cm}
\subfigure[]{\includegraphics[height= 0.2 \textheight,width=0.4 \textwidth]{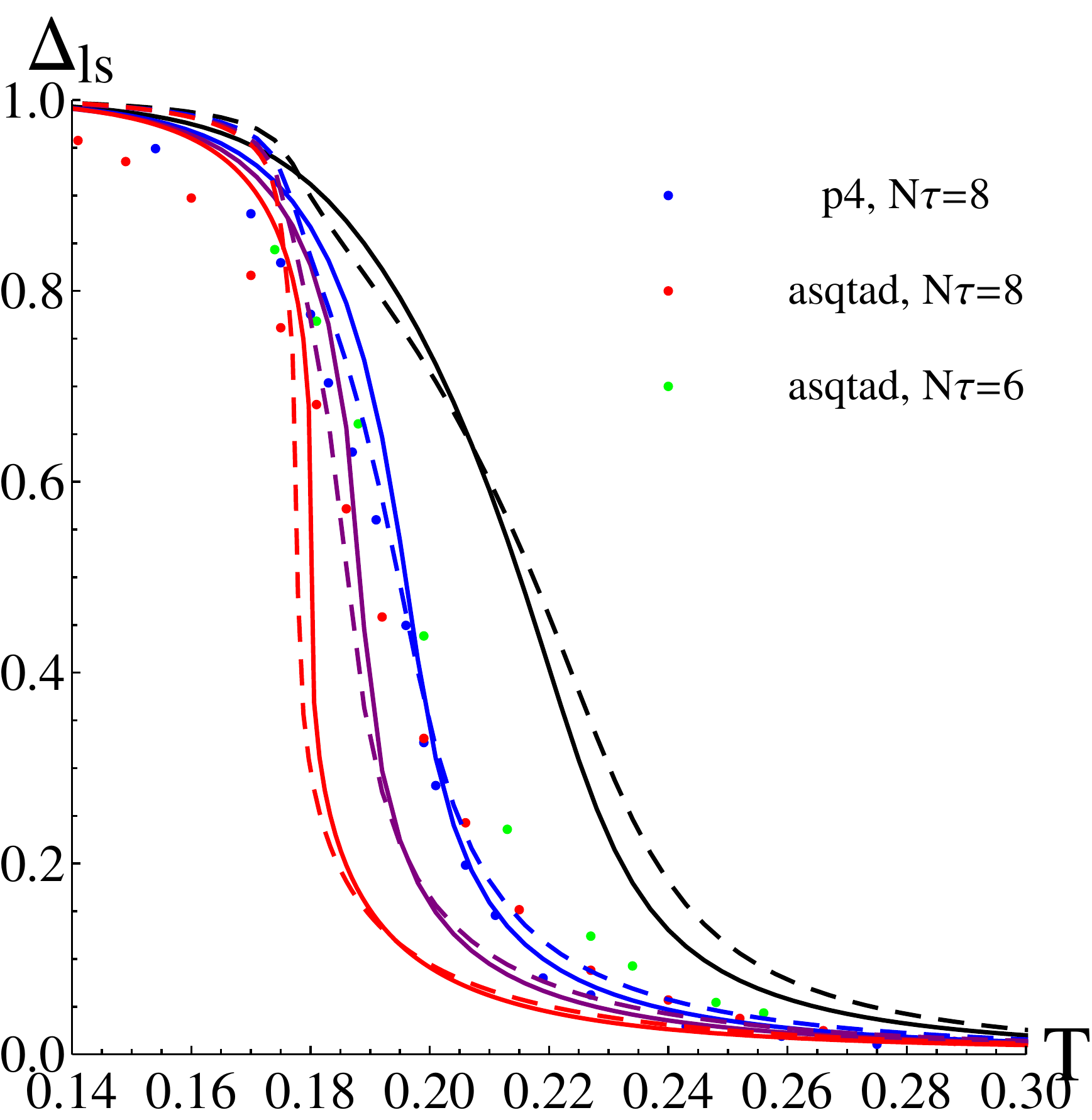}}
\caption{\footnotesize{Left: trace anomaly with $8q$ strength $g_1=6000$ GeV$^{-8}$, with ${\cal U_{P}}$ from \cite{Ratti:2006} and respective parameter $T_0=.19,.21,.23$ GeV (which yield peak positions from left to right). The lQCD data is taken from \cite{Bazavov:2009}. Right: the lQCD data for $\Delta_{ls}$, the subtracted chiral condensate value normalized to its zero T value, as defined in \cite{Bazavov:2009}, compared to PNJL calculations for several $g_1$ strengths, color code as in Fig. 2. Solid lines: with ${\cal U_{P}}$ from \cite{Ratti:2006}, dashed lines: with ${\cal U_{P}}$ from \cite{Roessner:2007}, both at $T_0=.19$ GeV.} }
\end{center}
\end{figure} 
\begin{figure}[!ht]
\begin{center}
\hspace*{-0.6cm}\includegraphics[height=2.5cm,width=5cm]{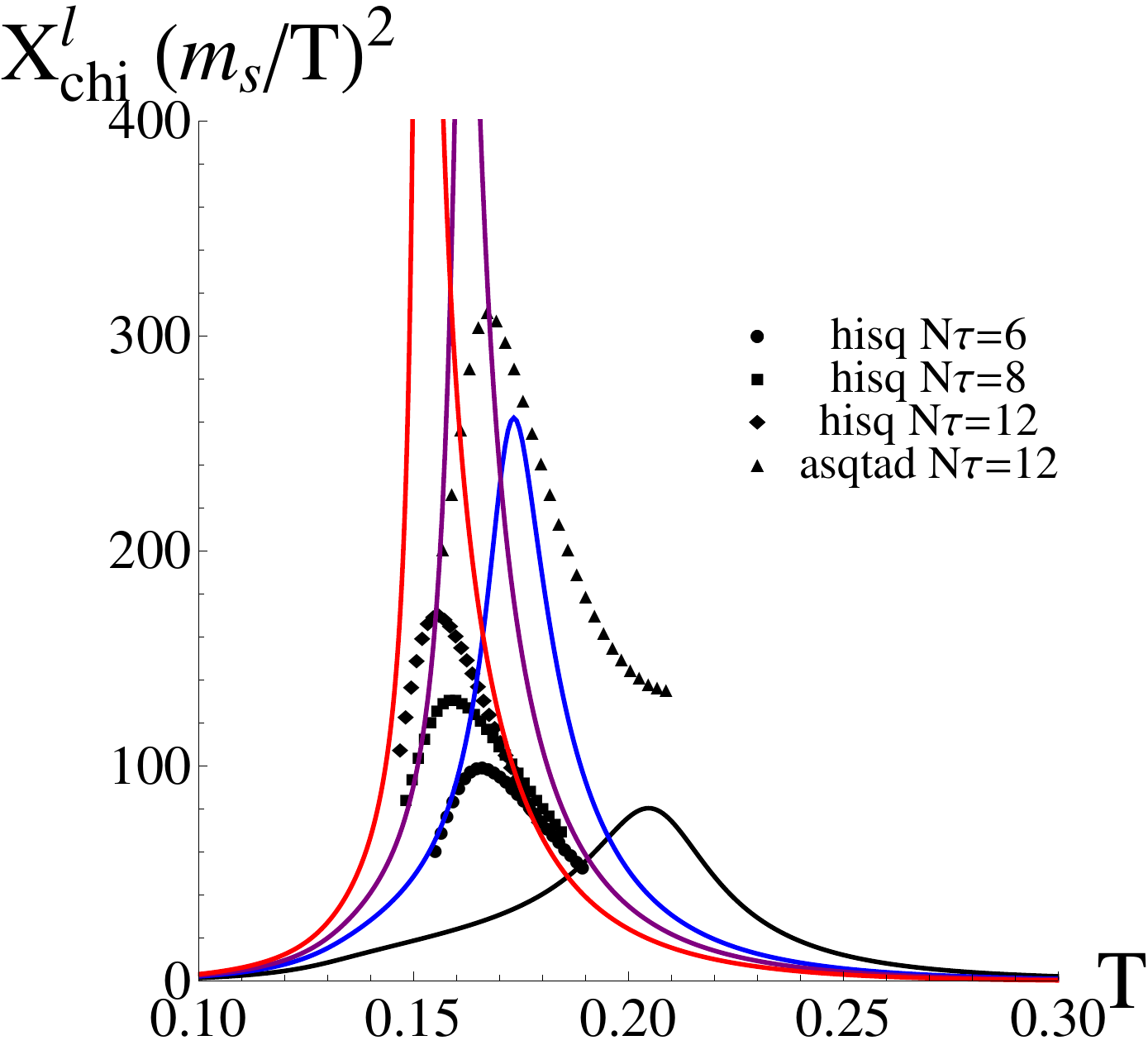}\hspace*{0.3cm}\includegraphics[height=2.5cm,width=5cm]{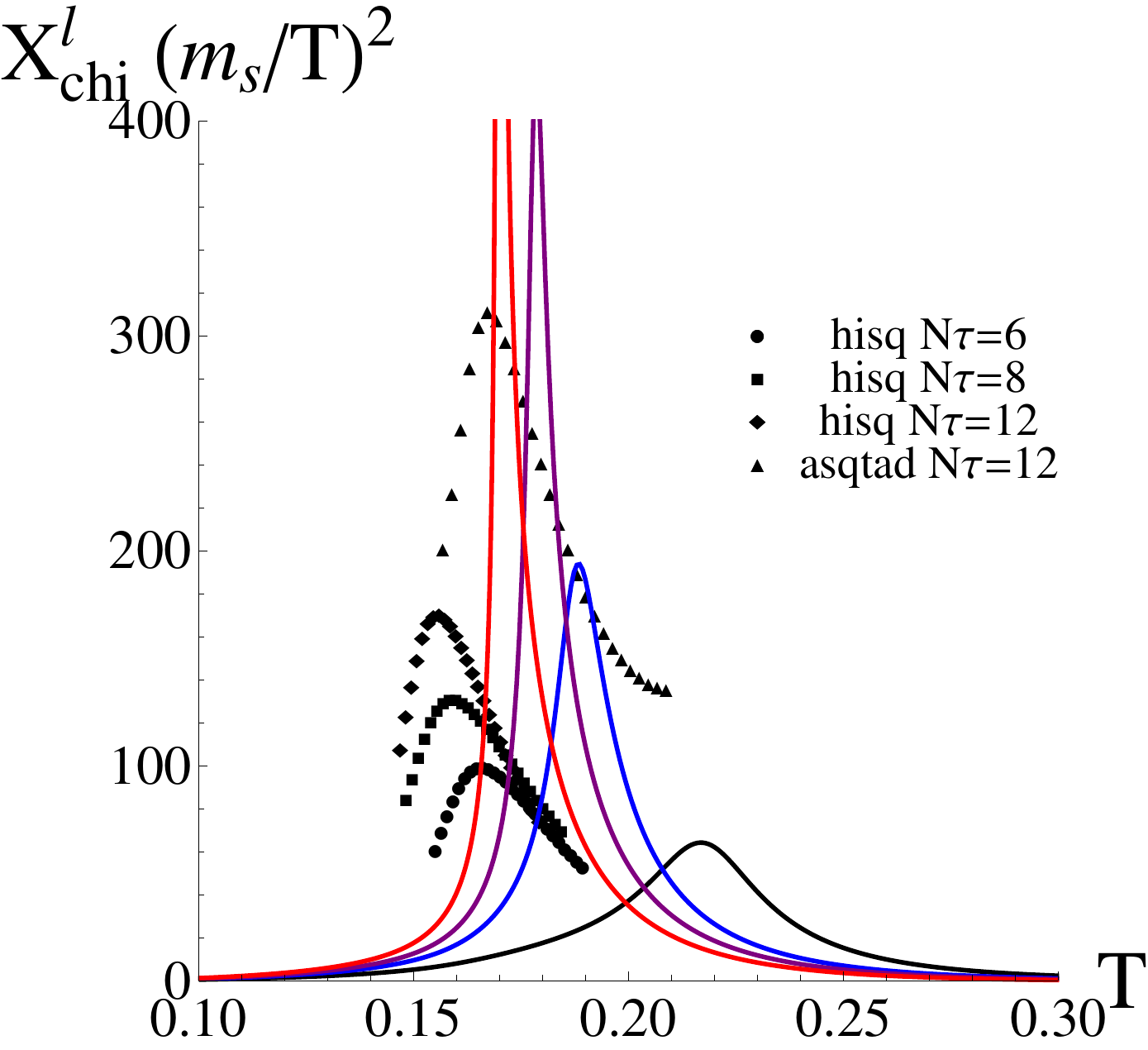}
\caption{\footnotesize The lQCD data for the light quark chiral susceptibility $X^{l}_{chi}$ in the continuum limit taken from \cite{Bazavov:2012}, in comparison with the PNJL model with ${\cal U_{P}}$ \cite{Ratti:2006} at $T_0=.15 GeV$ (left panel) and  $T_0=.19 GeV$ (right panel)  for different $g_1$ strengths (solid lines, narrower peaks correspond to increasing $g_1$).} 
\end{center}
\end{figure}

\end{document}